# The outburst of V713 Cephei in August 2009

David Boyd, Denis Denisenko, Robert Koff, Ian Miller, Bart Staels

**Abstract**

During the outburst of V713 Cephei in August 2009 the times of 8 eclipses were measured and these, together with 5 eclipse timings obtained during quiescence in August 2007, provide an improved orbital period of 0.085418432(4)d. No superhumps were observed in the light curve indicating this was a normal UG-type dwarf nova outburst. We found the eclipse depth decreased linearly with rising system excitation level, falling from ~ 3 magnitudes in quiescence to ~2 magnitudes during outburst. The depth and totality of eclipses in quiescence suggests a high orbital inclination. We saw no variation in the FWHM of eclipses between quiescence and outburst despite a significant change in shape of the eclipse profile.

## 1. Historical observations

The variability of V713 Cephei was originally discovered by S. V. Antipin on plates in the archive of the Sternberg Astronomical Insitute in Moscow (1). He found it in outburst at magnitude 15.5 on two plates taken in October and November 1989 and labelled it Var 75 Cep. An earlier outburst at magnitude 15.3 was subsequently discovered on two Sonneberg archive plates taken in August 1984. In 2006 it was officially assigned the name V713 Cep (2). The GCVS reports its maximum and minimum magnitudes as 15.3 and 18.8 (3). By virtue of its blue colour and apparent outburst amplitude of 3.5 magnitudes, it is classified in the GCVS as a UG-type dwarf nova.

The earliest observations of V713 Cep in the AAVSO database are in 2003, presumably following publication of the Antipin and Kroll paper (1). All positive observations recorded it around 18[th] magnitude or fainter apart from two, on 2008 June 25 at 15.03V and 2 days later at 16.98V. A search of the BAAVSS database produced no positive observations brighter than 18[th] magnitude.

## 2. Observations in 2007

It was discovered to be an eclipsing dwarf nova by Denisenko on 2007 August 25 (4) using the 1.5-m Russian Turkish Telescope. At that time it was in quiescence at around magnitude 18. Over the following 9 days, 5 eclipses about 3 magnitudes deep were recorded. An orbital period of 0.085419(2)d was calculated from these observations.

## 3. Observations in 2009

V713 Cep was reported rising to outburst on 2009 August 28.135 at magnitude 16.96C (5) and confirmed at mag 15.1C on August 28.795 (6). Observations by the authors between August 29.097 and August 30.136 recorded over 1200 images totalling 22 hours of exposure time. The equipment used is listed in Table 1. These images revealed a steadily fading light curve containing 8 eclipses, some of which were recorded by more than one observer. On August 31 V713 Cep was much fainter and on September 1 was measured at magnitude 19.1. According to the AAVSO database, in the 3 months prior to the outburst its mean recorded magnitude was 18.8 and the last observation before the start of the outburst was by Poyner on August 25.126 at 18.78C.

All images obtained during the 2009 outburst were dark-subtracted and flat-fielded and instrumental magnitudes measured by aperture photometry. Magnitudes for comparison stars were obtained by each observer from either the USNO-A2.0 or GSC 2.3 catalogues. The relative displacement in magnitude of the light curves between some observers which arose from the use of different comparison stars was removed by applying magnitude offsets to align the light curves for all observers. All times of observation were converted to Heliocentric Julian Dates (HJD). Figure 1 shows the resulting light curve of the 2009 outburst including the two initial observations noted above. The outburst appears to have peaked close to magnitude 15.0 making the outburst amplitude 3.8 magnitudes.

Astrometry on images during the 2009 outburst gives a mean position of V713 Cas as RA 20h 46m 38.67(1)s Dec 60° 38ø3.0(1)ö in good agreement with its position in USNO-B1.0.

**4. Orbital period**

The times and magnitudes at minimum of all the eclipses observed in 2007 and 2009 were obtained by quadratic fits to the lower part of each eclipse. In some cases coverage of the eclipse was relatively poor and this is reflected in larger errors on the times and magnitudes at minimum. By extrapolating separate linear fits to the 2007 and 2009 times of minimum, unambiguous cycle numbers covering both dates can be deduced. Using these cycle numbers, a weighted linear fit to the times of all 13 eclipses gives the following heliocentric ephemeris for the time of minimum:

$$HJD(min) = 2454338.37646(3) + 0.085418432(4) * E \qquad (1)$$

Table 2 lists the observed eclipse times of minimum, cycle numbers and observed minus calculated (O-C) values relative to the ephemeris in eqn (1) for the eclipses in 2007 and 2009. The O-C values are plotted in Figure 2.

**5. Orbital light curves**

Eclipses were removed from the 2009 light curve shown in Figure 1 and a 4$^{th}$ order polynomial fitted to the remaining data to establish a smooth underlying trend for the fading light curve. Figure 3 plots the residual light curve after removal of this underlying trend. Period analysis of this detrended light curve reveals a prominent signal at 0.0854(5)d consistent with the orbital period obtained above. Removing this period leaves no other significant signals. In particular, there is no evidence of a superhump signal so it appears that this was a normal UG-type outburst.

Figures 4 and 5 show the 2007 light curve and the 2009 detrended light curve respectively, both phased at the orbital period with phase zero at the eclipse minimum. In quiescence (Figure 4) a ~0.3 magnitude orbital hump is visible preceding the eclipse. This is caused by the passage across the front of the accretion disk of the bright spot where gas streaming from the secondary star impacts the edge of the disk. During outburst (Figure 5) the orbital hump is no longer visible as the bright spot is now swamped by the increased luminosity of the accretion disk.

**6. Eclipse profiles**

Figure 6 shows the 2007 eclipse light curves superimposed and aligned on their times of minimum and the out-of-eclipse magnitude level. With a quiescent eclipse depth of ~3 magnitudes, this is one of the most deeply eclipsing dwarf novae. The shape of the superimposed eclipses, although poorly sampled, is reminiscent of that observed in other dwarf novae in quiescence, for example Z Cha (7) and IY UMa (8). The eclipse ingress takes place in two stages with the white dwarf disappearing first followed a short time later by the bright spot. It is tempting the see the initial drop of ~1.8 magnitudes as the disappearance of the white dwarf. This is very rapid because of the relatively small size of the white dwarf. There is then a brief shoulder and a further drop of just over a magnitude as the bright spot is eclipsed. The bottom of the eclipse is approximately flat indicating that the white dwarf and bright spot are totally eclipsed and that the binary system is therefore likely to have a high orbital inclination. The end of the eclipse is initiated by emergence of the white dwarf. As the eclipse egress is not well observed, we don¢t detect the separate emergence of the bright spot. The accretion disk, being relatively small and faint in quiescence, is likely to make little contribution to the eclipse profile.

Figure 7 shows superimposed eclipses from two nights during 2009 relative to the detrended light curve described above. The eclipse profile in outburst is noticeably different from that in quiescence being V-shaped, less deep and slightly broader at the top. In outburst the bright, expanded accretion disk dominates the light output of the system and the V-shape of the eclipse suggests the expanded disk is not completely eclipsed by the secondary. The white dwarf and bright spot are out-shone by the disk during outburst and make relatively little contribution to the total light.

To examine the variation in eclipse parameters with different excitation states of the system, we adopt the mean out-of-eclipse magnitude as a measure of the system excitation level. The depth of each eclipse during outburst was measured by taking the difference in magnitude between the $4^{th}$ order polynomial trend at mid-eclipse and the fitted minimum magnitude. In quiescence we used the mean out-of-eclipse magnitude excluding the orbital hump. The variation of eclipse depth with out-of-eclipse magnitude is shown in Figure 8. There is an apparently linear trend of decreasing eclipse depth with rising system excitation level. Similar results have been found for the eclipsing dwarf nova SDSS J150240.98+333423.9 (9).

We also measured the full width at half maximum (FWHM) of the three superimposed eclipse profiles in Figures 6 and 7 by fitting straight lines to the approximately linear segments of both sides of the eclipse profiles. Figure 9 shows that there is little variation of eclipse width with system excitation level. This is similar to the behaviour of the eclipsing UGSS dwarf nova CG Dra (10) and contrasts with eclipsing UGSU systems where evidence suggests eclipses double in width during superoutbursts (8, 9, 11). This is consistent with the accepted interpretation that in a superoutburst the disk expands to a greater extent than in a normal outburst and sufficiently so to excite a 3:1 resonance with the orbit of the secondary.

**7. Future observations**

This was the first well-observed outburst of V713 Cep. If the duration and magnitude of this outburst is typical, it is very likely that other outbursts have been missed as it is too faint in outburst to be reliably detected visually. Indeed Poyner observed the star visually on 2008 June 25, the same night it was detected by CDD at 15.03V, but could only confirm that it was fainter than mag 14.8. It is therefore worth monitoring V713 Cep for further outbursts by CCD photometry.

**8. Acknowledgements**

We acknowledge with thanks variable star observations from the AAVSO International Database contributed by observers worldwide and used in this research. We thank Clive Beech for his assistance in providing information from the BAAVSS Database. We are also grateful for use of the Simbad and Vizier services operated by CDS Strasbourg and to the referee Chris Lloyd for helpful comments.

**Addresses**

DB:  5 Silver Lane, West Challow, Wantage, Oxon, OX12 9TX, UK [drsboyd@dsl.pipex.com]
DD:  Space Research Institute (IKI), Russian Academy of Sciences, Moscow, Russia
      [denis@hea.iki.rssi.ru]

RK:  CBA Colorado, 980 Antelope Drive West, Bennett, CO 80102, USA [bob@antelopehillobservatory.org]
IM:  Furzehill House, Ilston, Swansea, SA2 7LE, UK [furzehillobservatory@hotmail.com]
BS:  CBA Flanders (Patrick Mergan Observatory), Koningshofbaan 51, B-9308 Hofstade, Belgium [staels.bart.bvba@pandora.be]

| Observer | Equipment |
|---|---|
| DB | 0.35-m f/5.3 SCT + SXV-H9 CCD |
| RK | 0.25-m f/10 SCT + AP47 CCD |
| IM | 0.35-m f/10 SCT + SXV-H16 CCD |
| BS | 0.28-m f/6.3 SCT + MX716 CCD |

Table 1: Equipment used to observe the 2009 outburst.

| Date | Time of minimum (HJD) | Error (d) | Cycle no | O-C (d) | ID |
|---|---|---|---|---|---|
| 2007-Aug-25 | 2454338.37651 | 0.00008 | 0 | 0.00005 | DD1 |
| 2007-Aug-25 | 2454338.46181 | 0.00007 | 1 | -0.00007 | DD2 |
| 2007-Aug-26 | 2454339.48692 | 0.00004 | 13 | 0.00002 | DD3 |
| 2007-Aug-29 | 2454342.39093 | 0.00013 | 47 | -0.00019 | DD4 |
| 2007-Sep-03 | 2454347.43104 | 0.00021 | 106 | 0.00023 | DD5 |
| 2009-Aug-29 | 2455072.63328 | 0.00007 | 8596 | -0.00002 | RK1 |
| 2009-Aug-29 | 2455072.71861 | 0.00007 | 8597 | -0.00011 | RK2 |
| 2009-Aug-29 | 2455072.80417 | 0.00003 | 8598 | 0.00004 | RK3 |
| 2009-Aug-29 | 2455072.88953 | 0.00004 | 8599 | -0.00002 | RK4 |
| 2009-Aug-29 | 2455072.97504 | 0.00006 | 8600 | 0.00006 | RK5 |
| 2009-Aug-29 | 2455073.40204 | 0.00002 | 8605 | -0.00003 | BS1 |
| 2009-Aug-29 | 2455073.40207 | 0.00002 | 8605 | 0.00001 | DB1 |
| 2009-Aug-29 | 2455073.40220 | 0.00003 | 8605 | 0.00013 | IM |
| 2009-Aug-29 | 2455073.48742 | 0.00010 | 8606 | -0.00007 | BS2 |
| 2009-Aug-29 | 2455073.48743 | 0.00003 | 8606 | -0.00006 | DB2 |
| 2009-Aug-30 | 2455073.57282 | 0.00013 | 8607 | -0.00008 | BS3 |

Table 2. Eclipse times of minimum and O-C values with respect to the ephemeris in eqn (1).

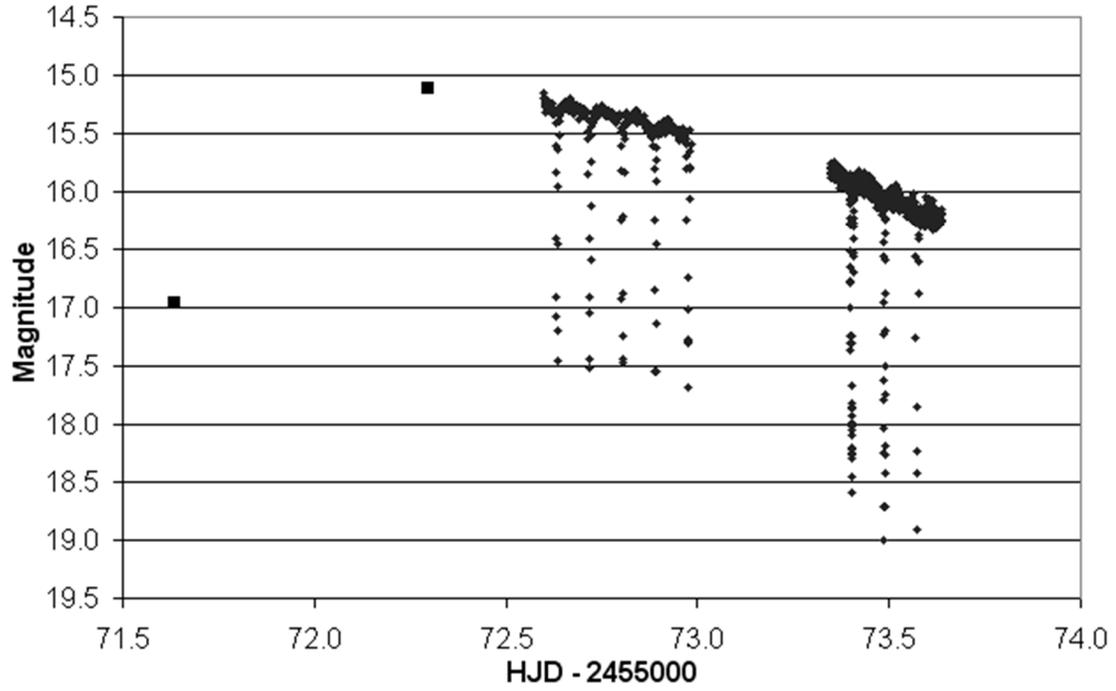

Figure 1. Light curve of the 2009 outburst.

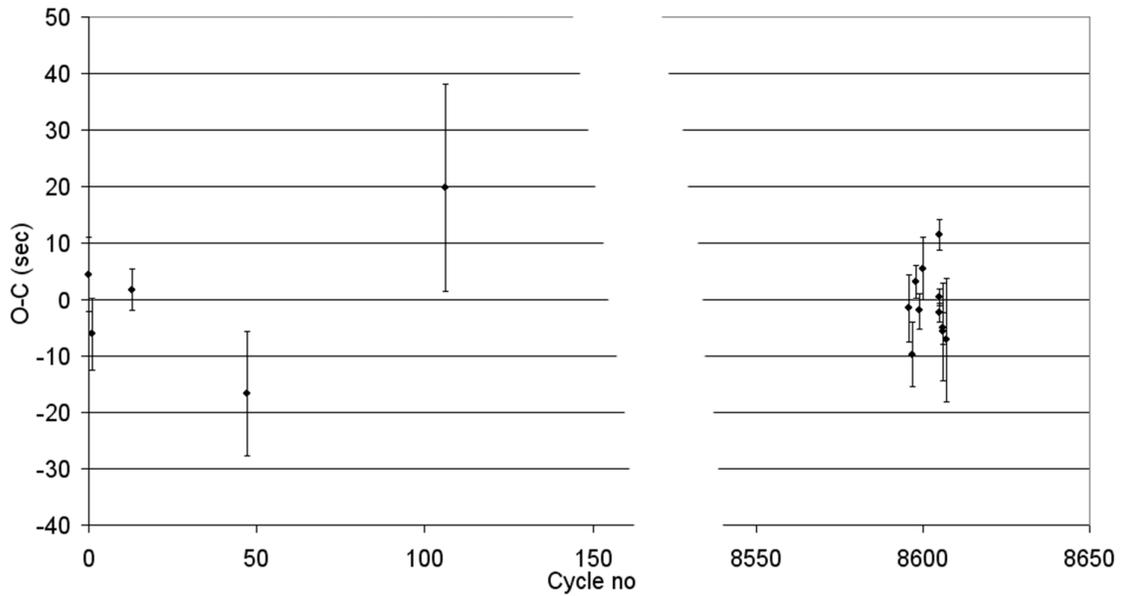

Figure 2. O-C values for all eclipse times of minimum with respect to the ephemeris in eqn (1).

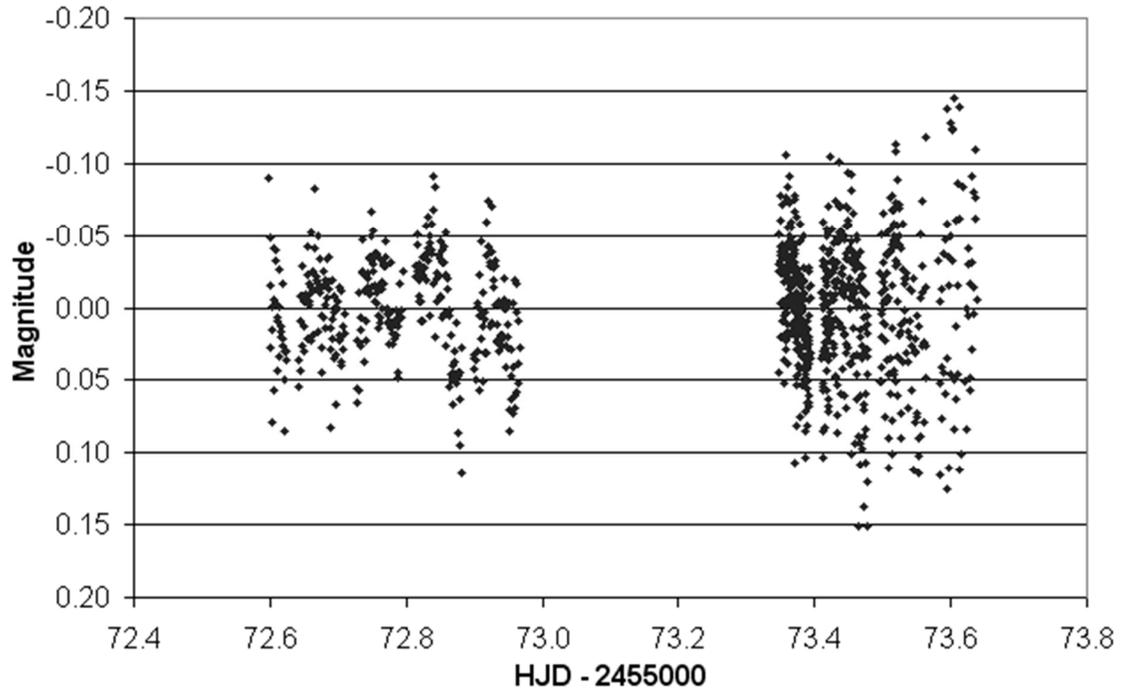

Figure 3. Residuals of the 2009 magnitude measurements outside eclipse with respect to the underlying trend.

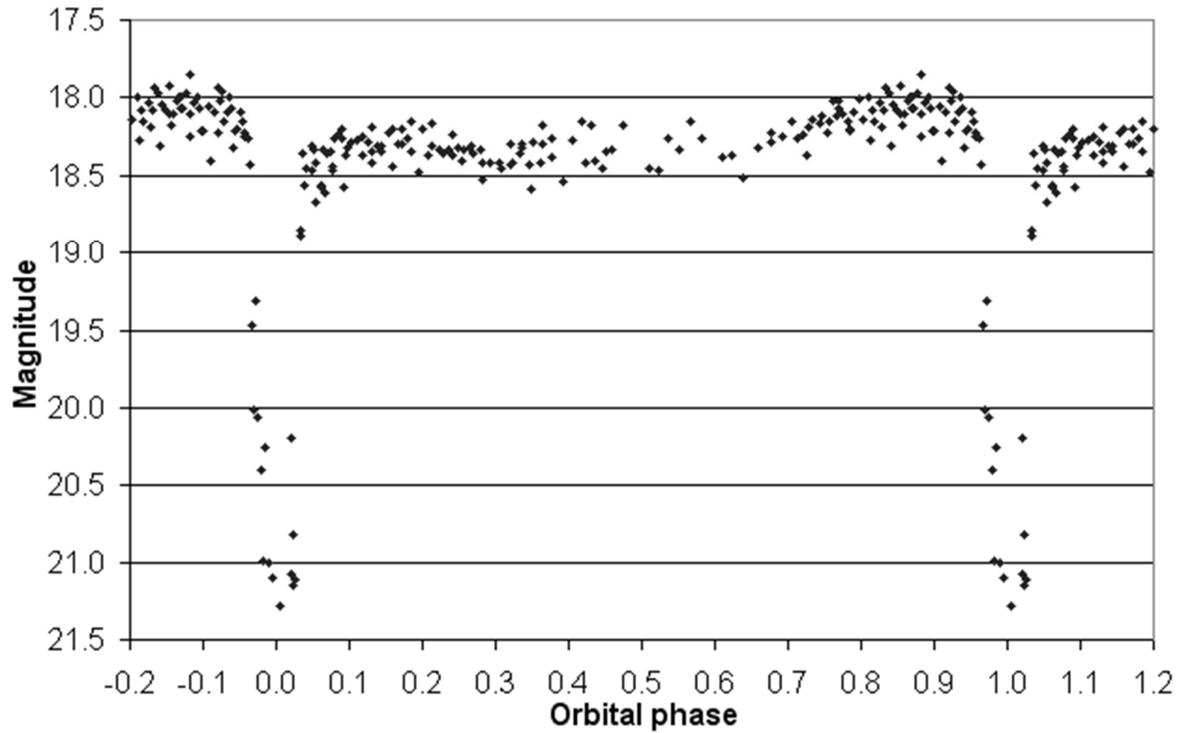

Figure 4. Observations in quiescence in 2007 phased at the orbital period.

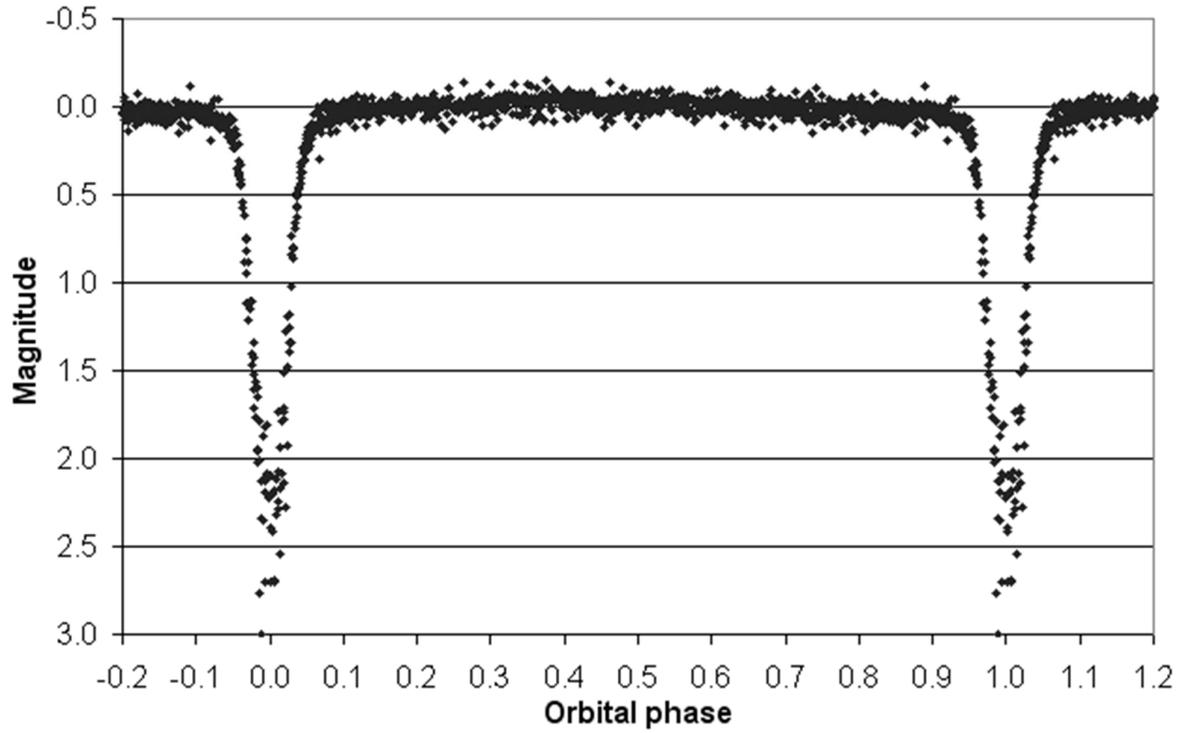

Figure 5. Observations in outburst in 2009 after removal of the underlying trend and phased at the orbital period.

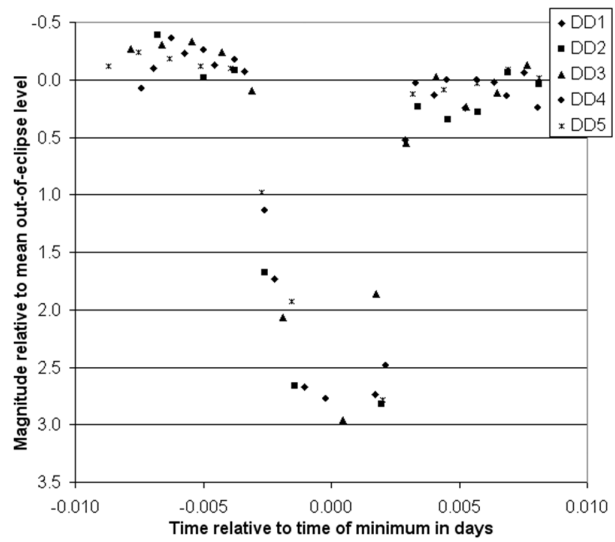

Figure 6. Superimposed eclipses in quiescence in 2007.

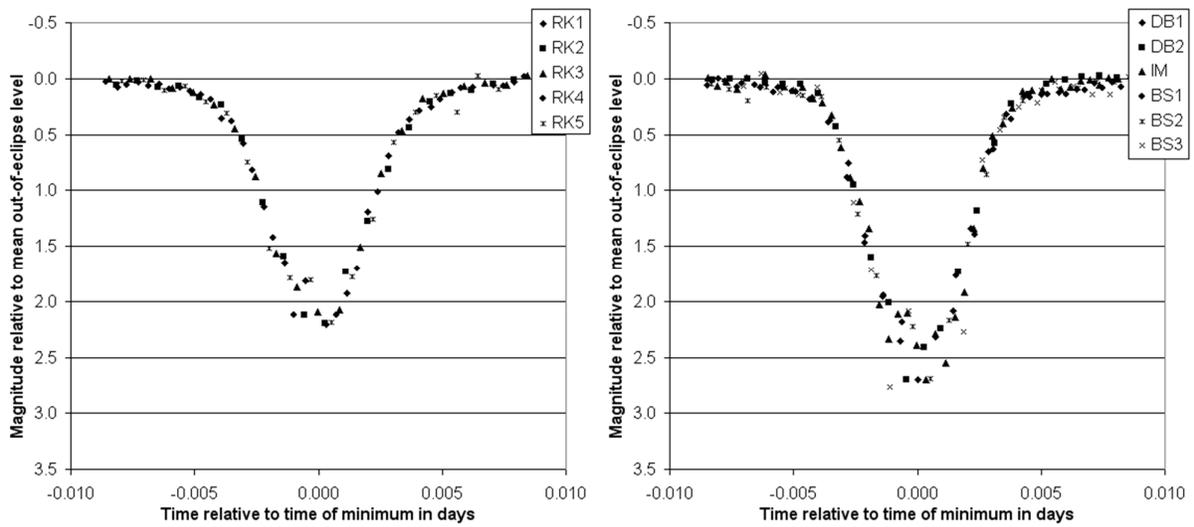

Figure 7. Superimposed eclipses during outburst on 2009 Aug 29 - 30.

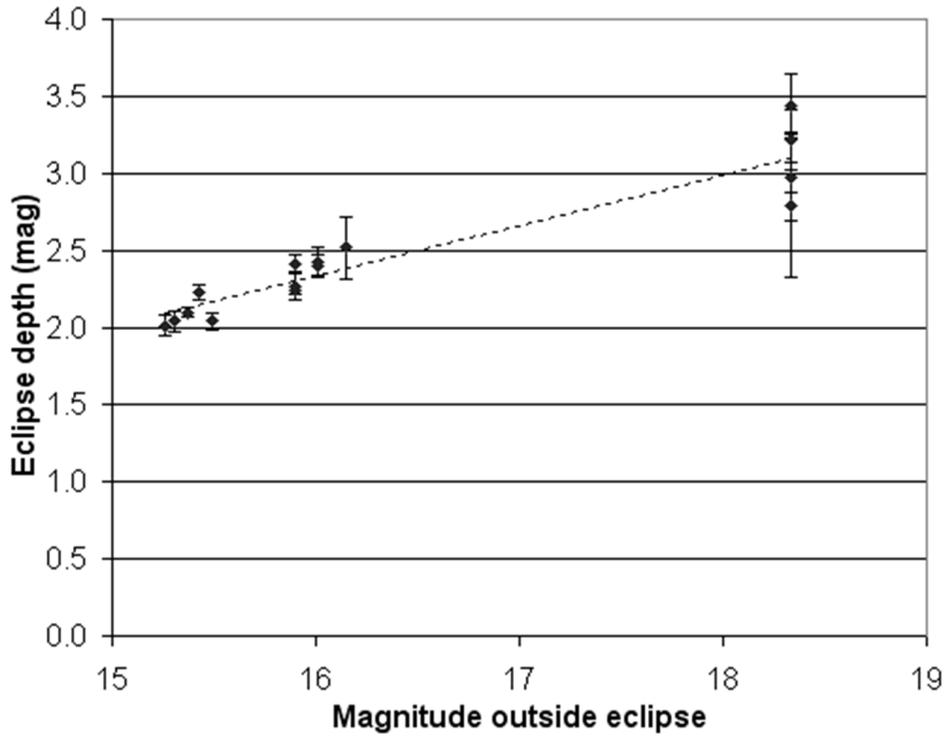

Figure 8. Variation of eclipse depth with magnitude outside eclipse.

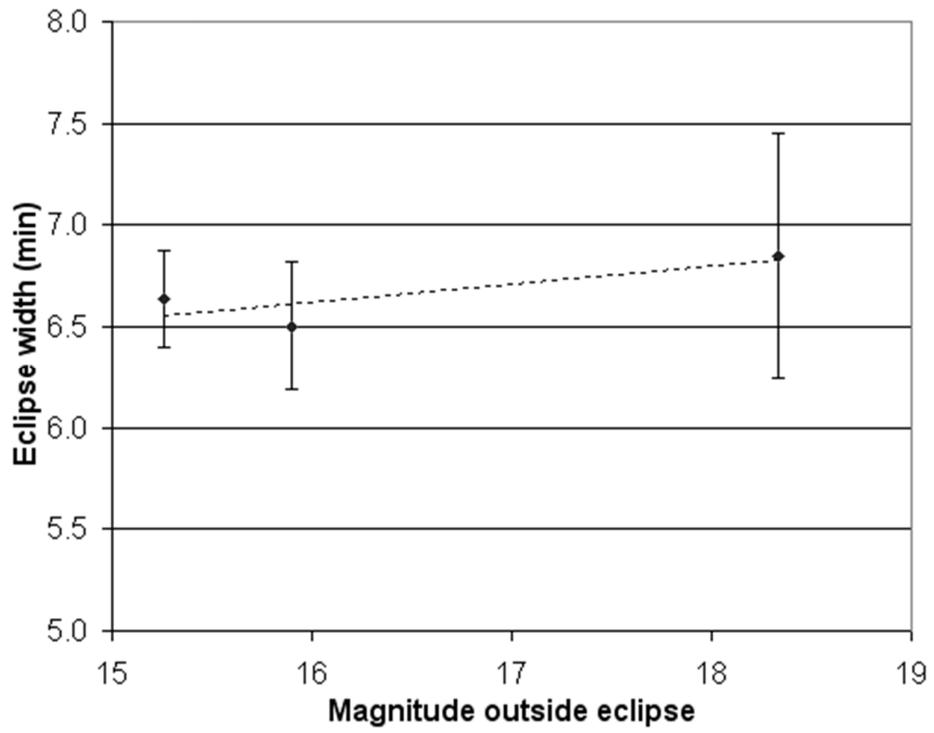

Figure 9. Variation of eclipse width with magnitude outside eclipse.